\documentclass{article}
\usepackage{spconf}

\usepackage{color,xcolor}
\usepackage{epsfig}
\usepackage{graphicx}
\usepackage{lipsum}
\usepackage{pifont}
\usepackage{array}
\usepackage{booktabs}
\usepackage{colortbl}
\usepackage{multirow}
\usepackage{float}
\usepackage{caption}
\usepackage{adjustbox}
\usepackage{subfigure}
\usepackage{footnote}
\makesavenoteenv{tabular}
\makesavenoteenv{table}

\usepackage{amsmath,amsfonts,amssymb,bm}
\usepackage[super]{nth}

\usepackage{changepage}
\usepackage{extramarks}
\usepackage{lastpage}
\usepackage{setspace}
\usepackage{soul}
\usepackage{xspace}

\usepackage{url}
\usepackage{hyperref}
\hypersetup{colorlinks=True, urlcolor=black}

\usepackage{algorithm, algorithmic}
\usepackage{enumitem}
\usepackage{verbatim}
\usepackage{pifont}
\usepackage[acronyms]{glossaries}
\glsdisablehyper

\newcommand{\ssectdot}[1]{Sec.~\ref{ssec:#1}}

\newcommand{\figdot}[1]{Fig.~\ref{fig:#1}}
\newcommand{\tbl}[1]{Table~\ref{tab:#1}}

\newcommand{\ignore}[1]{}

\makeatletter
\DeclareRobustCommand\onedot{\futurelet\@let@token\@onedot}
\def\@onedot{\ifx\@let@token.\else.\null\fi\xspace}

\makeatother

\definecolor{MyDarkBlue}{rgb}{0,0.08,1}
\definecolor{MyDarkGreen}{rgb}{0.02,0.6,0.02}
\definecolor{MyDarkRed}{rgb}{0.8,0.02,0.02}
\definecolor{MyDarkOrange}{rgb}{0.40,0.2,0.02}
\definecolor{MyPurple}{RGB}{111,0,255}
\definecolor{MyRed}{rgb}{1.0,0.0,0.0}
\definecolor{MyGold}{rgb}{0.75,0.6,0.12}
\definecolor{MyDarkgray}{rgb}{0.66, 0.66, 0.66}

\usepackage{pgfplots}
\DeclareUnicodeCharacter{2212}{−}
\usepgfplotslibrary{groupplots,dateplot}
\usetikzlibrary{patterns,shapes.arrows}
\pgfplotsset{compat=newest}

\title{Towards Universal Speech Discrete Tokens: A Case Study for ASR and TTS}

\name{
    Yifan Yang$^{\star,1}$,
    Feiyu Shen$^{\star,1}$,
    Chenpeng Du$^{1}$,
    Ziyang Ma$^{1}$,
    Kai Yu$^{1}$,
    Daniel Povey$^{\dag,2}$,
    Xie Chen$^{\dag,1}$
    \thanks{$\star$~stands for equal contribution. $\dag$~stands for corresponding authors.\\
    This work was supported by the National Natural Science Foundation of China (No. 62206171), and the International Cooperation Project of PCL, and in part by Shanghai Municipal Science and Technology Major Project under Grant 2021SHZDZX0102.
    }
}

\address{
    $^{1}$ MoE Key Lab of Artificial Intelligence, AI Institute \\
    X-LANCE Lab, Department of Computer Science and Engineering, Shanghai Jiao Tong University, China \\
    $^{2}$ Xiaomi Corporation, Beijing, China
}

\begin{document}
\ninept

\maketitle

\begin{abstract}
Self-supervised learning (SSL) proficiency in speech-related tasks has driven research into utilizing discrete tokens for speech tasks like recognition and translation, which offer lower storage requirements and great potential to employ natural language processing techniques. However, these studies, mainly single-task focused, faced challenges like overfitting and performance degradation in speech recognition tasks, often at the cost of sacrificing performance in multi-task scenarios. This study presents a comprehensive comparison and optimization of discrete tokens generated by various leading SSL models in speech recognition and synthesis tasks. We aim to explore the universality of speech discrete tokens across multiple speech tasks. Experimental results demonstrate that discrete tokens achieve comparable results against systems trained on FBank features in speech recognition tasks and outperform mel-spectrogram features in speech synthesis in subjective and objective metrics. These findings suggest that universal discrete tokens have enormous potential in various speech-related tasks. Our work is open-source and publicly available at https://github.com/k2-fsa/icefall.
\end{abstract}
\vspace{-1em}
\section{Introduction}
\vspace{-0.5em}
Self-supervised learning (SSL) \cite{SSL, wav2vec2, vq-wav2vec, HuBERT, WavLM, ma2022mt4ssl, ma2023pushing} is a robust methodology capable of leveraging large amounts of unlabeled data to deliver substantial performance improvements. In speech-related downstream tasks, representations derived from SSL stand out for their generalizability and accessibility.
While most existing research focuses on utilizing real-valued features from intermediate layers in SSL models for downstream tasks, there is scope for investigating the use of discretized token sequences.
Speech discrete tokens can be broadly divided into semantic tokens and acoustic tokens \cite{AudioLM}. Models including vq-wav2vec~\cite{vq-wav2vec}, wav2vec 2.0~\cite{wav2vec2}, HuBERT~\cite{HuBERT}, and WavLM~\cite{WavLM} generate semantic tokens trained for discrimination or masking prediction, maintaining linguistic information of the speech. In contrast, acoustic tokens, produced by audio neural codec models like Soundstream~\cite{Soundstream} and Encodec~\cite{EnCodec}, aim to reconstruct speech accurately, reflecting acoustic details.

Previous works have studied the application of discrete tokens in individual speech tasks.
Baevski et al.~\cite{baevski_discreteASR} examined the utilization of discrete tokens in automatic speech recognition (ASR) and found underperformed compared to traditional FBank features unless combined with a BERT~\cite{BERT} as a preprocessing step.
Chang et al.~\cite{cmu_discreteASR} employed refined discretized token sequences via de-duplication and subword modeling techniques, approximating the performance of conventional acoustic features while reducing computational and storage costs.
Du et al.~\cite{VQTTS} presented VQTTS, incorporating a txt2vec acoustic model and a vec2wav vocoder. This system utilizes self-supervised Vector Quantization (VQ) acoustic features to mitigate the adverse impact of the disparity between the ground truth and predicted mel-spectrograms, thereby enhancing the overall text-to-speech (TTS) system performance.
Subsequently, Du et al.~\cite{UniCATS} introduced UniCATS, a TTS system based on semantic tokens and free of acoustic tokens and speaker embeddings. It integrates both semantic and acoustic contexts through the use of contextual VQ-diffusion in its CTX-txt2vec component and vocoding in its CTX-vec2wav component, achieving state-of-the-art (SOTA) TTS performance. 
Nguyen et al.~\cite{EXPRESSO} utilized discrete tokens from HuBERT and Encodec for speech resynthesis and found Encodec, designed for general audio compression, outperformed HuBERT units in this task.

The application of discrete tokens in multiple tasks remains preliminary.
Recently, Wang et al.~\cite{VIOLA} presented VioLA, transforming speech signals into discrete tokens, enabling a unified conditional language modeling way for multiple spoken processing tasks. 
However, they only employ codec to build a multi-task model with limited performance compared to single tasks. Systematic experiments on different tasks using varied discrete tokens are still lacking.

\begin{figure}
    \centering
    \includegraphics[scale=0.25]{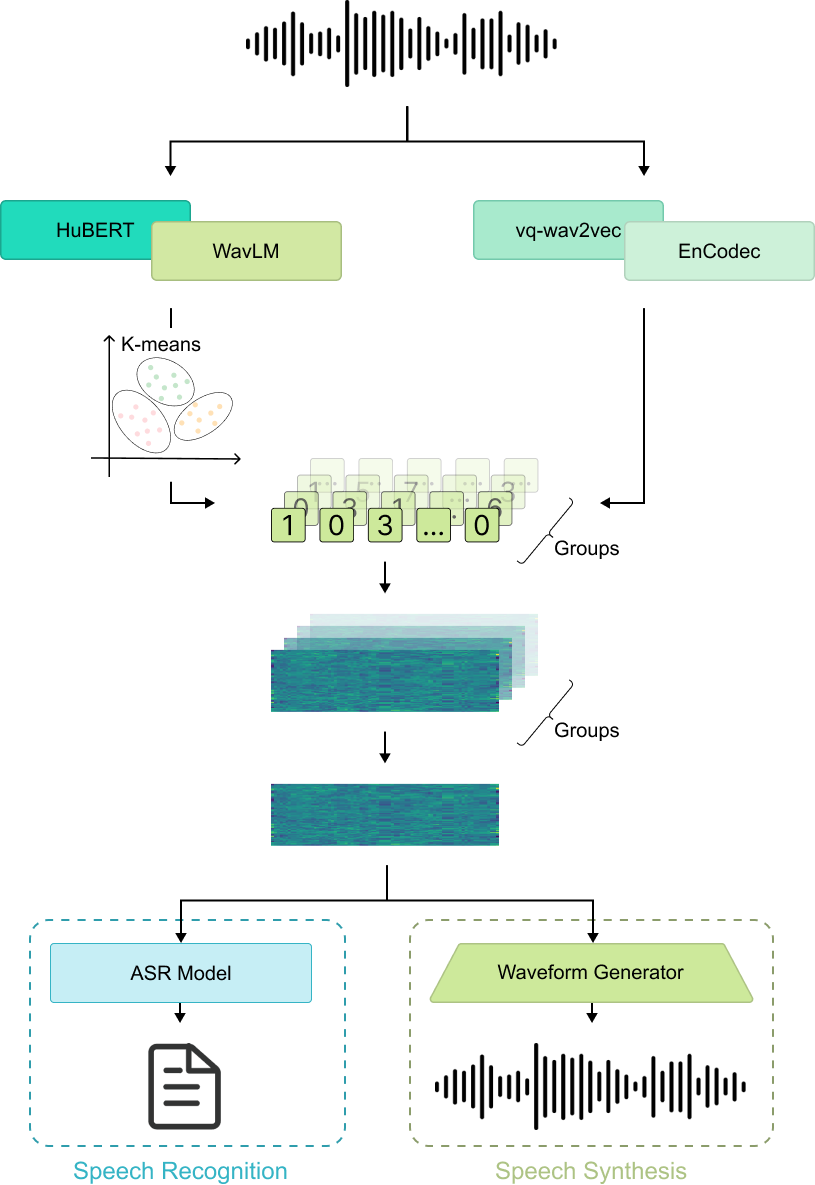}
    \caption{Illustration of the pipeline for speech discrete tokens.}
    \label{fig:System}
    \vspace{-1.6em}
\end{figure}

In this context, this study aims to investigate the universality of speech discrete tokens across multiple speech-related tasks. We take two representative speech processing tasks, speech recognition and speech synthesis, for a case study.
As illustrated in \figdot{System}, we transform raw audio into speech discrete tokens using four leading SSL models, including vq-wav2vec, encodec, HuBERT, and WavLM.
For ASR, we utilize these discrete tokens and their associated texts to train end-to-end (E2E) ASR models~\cite{li2022recent}. We introduce customized data augmentation techniques for discretized inputs to mitigate overfitting. In order to get a whole picture of the potential of discrete tokens, we train and evaluate this ASR model on three distinct data sets, including LibriSpeech, GigaSpeech, and AISHELL-1. Note that AISHELL-1 is a Mandarin speech corpus, while all SSL models are pretrained with English audio.
For TTS, we conduct experiments on speech resynthesis tasks to explore its upper bound without interference from the acoustic model. we trained vocoders with various discrete tokens on LibriTTS and compared their performance through both subjective and objective evaluations. Notably, using discrete tokens yields superior performance compared to conventional real-valued mel-spectrogram features.
\vspace{-0.8em}
\section{Discrete Speech Tokens from SSL Models}
\vspace{-0.5em}
In self-supervised learning models, continuous audio can be discretized using two primary methods: VQ and k-means clustering on embeddings.
vq-wav2vec employs autoregressive Convolutional Neural Networks (CNNs) to extract feature representations from raw audio, which are then organized into multiple groups. Subsequently, vector quantization is applied in each group, facilitated by either the Gumbel-Softmax or online k-means clustering, yielding multiple sets of discrete token sequences.
By contrast, EnCodec applies Residual Vector Quantization (RVQ) \cite{Soundstream} on the output of the convolutional-based encoder.
For HuBERT and WavLM, the discretization of speech can be realized by applying k-means clustering to the hidden embeddings from some specified Transformer-Encoder layer.

In practice, a pre-trained vq-wav2vec model is employed to process 16 kHz audio and outputs embeddings at a 100 Hz rate with 2 groups. A pre-trained EnCodec processes 24 kHz audio at varying bitrates, which generates 75 Hz embeddings from 24 kHz inputs. For RVQ, eight hierarchical quantizers with 1024 entries are employed.
Hidden embeddings are extracted using pre-trained WavLM Large and HuBERT Large models. The final Transformer-Encoder layer is chosen for the ASR task according to the semantic correlation analysis in~\cite{superb}.

Discretization granularity can be adjusted by varying the number of clusters in the k-means algorithm. Specifically, 2000 clusters are chosen based on reports that more discrete tokens boost phone-normalized mutual information (PNMI) and enhance ASR performance~\cite{cmu_discreteASR}.

\vspace{-0.5em}
\section{Speech Recognition with Discrete Tokens}
\subsection{ASR with Discretized Input}
We individually employ discrete speech tokens of each type, along with their corresponding texts, to train distinct E2E ASR models, leveraging RNN-T loss for optimization.
These tokens are projected to 80 dimensions through a linear embedding layer. When several feature groups exist, each one is embedded, concatenated, and then projected to match the length of a single group. Subsequently, these features are interpolated to a uniform 100 Hz rate before feeding into the ASR model.

\vspace{-1.1em}
\subsection{Discretized Input Augmentation Policy}
\vspace{-0.5em}
\label{ssec:DiscretizedInputAugmentationPolicy}
We engineer data augmentation techniques tailored for discretized inputs to mitigate overfitting and enhance robustness for discrete token representations.
Inspired by the SpecAugment~\cite{SpecAugment} strategy for continuous features like FBanks, our policy incorporates the following deformations on the discrete tokens sequence $\boldsymbol{x} = \left( x_{1}, \cdots, x_{T} \right)$ of length $T$, and its corresponding embeddings sequence $\boldsymbol{f} = \left(f_{1}, \cdots, f_T\right)$:

\noindent\textbf{Time Warping} is implemented using the \texttt{interpolate} function of \texttt{PyTorch} in \texttt{nearest} mode. A time warp factor $W = 80$ is employed. A random centre $C$ is selected from $\left[ W + 1, T - W \right)$, and resultant warped size $S$ is determined within $\left[ C - W, C + W + 1 \right)$. Consequently, the sequence warps at the segments $\boldsymbol{x_{1:C-1}}$ and $\boldsymbol{x_{C:T}}$ to size $S$ and $T - S$, respectively.

\noindent\textbf{Time Masking} is executed on consecutive temporal steps of $\boldsymbol{x}$. $N$ denotes the number of mask regions, limited to $\min \left ( 10, \left \lceil 0.0015 \times T \right \rceil \right)$. The maximum mask width $M$ is set as $\min \left( 100, \left \lfloor \frac{0.15 \times T}{N} \right \rfloor \right)$. Each mask $i$ has a random width $m_i$ up to $M$ and starts at the time step $\lambda (T - m_i) $, where $\lambda \sim{\mathcal U[0, 1)}$.

\noindent\textbf{Embedding Masking} is performed successively along the vertical embedding dimension of $\boldsymbol{f}$. The masking stride is determined by a random value $m_j$ ranging from 0 to 27, and the starting position of the embedding masking is $\lambda (F - m_i) $, where $\lambda \sim{\mathcal U[0, 1)}$ and $F$ is the embedding dimension. The embedding masking is applied twice independently to improve the robustness.

\noindent\textbf{Gaussian Noise} followed by standard normal distribution is added to $\boldsymbol{f}$ with a probability of 0.25.

In the same batch, each audio sample is independently augmented with a probability of 0.9. All masking values are set to zero.
\vspace{-2em}
\section{Speech Synthesis with Discrete Tokens}
\vspace{-0.5em}
To investigate TTS performance among various discrete tokens, we conduct experiments on speech resynthesis tasks as it eliminates possible interference from the acoustic model and serves as an upper bound of TTS systems. In this study, we utilize CTX-vec2wav proposed in UniCATS~\cite{UniCATS}, which has demonstrated SOTA performance among vocoders operating on discrete tokens. Since discrete tokens primarily contain speech content, additional information, such as speaker identity or prosody, must be incorporated for resynthesis. In contrast to other approaches that utilize explicit speaker embedding or acoustic token prefixes, CTX-vec2wav employs contextual vocoding by utilizing mel-spectrogram as the prompt.

The CTX-vec2wav consists of a frontend encoding discrete tokens and a HifiGAN~\cite{HiFi-GAN} generator. The semantic tokens are first projected and encoded by two Conformer~\cite{Conformer} encoders with an auxiliary feature adapter in between. The auxiliary feature adapter, similar to the variance predictor in FastSpeech2~\cite{fastspeech2}, incorporates a three-dimensional prosody feature (pitch, energy, probability-of-voiced) to enhance audio quality. To condition on the mel-spectrogram prompt, the two Conformer encoders are equipped with an extra cross-attention layer following self-attention, which takes convolutional encoded mel-spectrogram as key and value. Then, the output from the second Conformer encoder is fed to HifiGAN to produce waveform.
\vspace{-1em}
\section{Experiments}
\vspace{-0.5em}
\subsection{ASR Experimental Setups}
The efficacy of the proposed protocol in ASR is assessed on LibriSpeech~\cite{LibriSpeech}, GigaSpeech~\cite{GigaSpeech}, and AISHELL-1~\cite{aishell1} corpora. 
Performance metrics include Word Error Rate (WER) for English and Character Error Rate (CER) for Chinese, assessed on LibriSpeech's test-clean and test-other sets, GigaSpeech's development and test sets, and AISHELL-1's development and test sets.
Separate k-means models are trained for each data set on a 100-hour, randomly selected subset of speech.

In FBank-based experiments, SpecAugment~\cite{SpecAugment} is applied during training for robustness. The input is 80-channel FBank features extracted over windows of 25ms strided by 10ms. The classification units are 500-class Byte Pair Encoding (BPE)~\cite{bpe} word pieces. In discrete tokens experiments, training incorporates discretized input augmentations as described in \ssectdot{DiscretizedInputAugmentationPolicy}.

The neural Transducer architecture is adopted for ASR. Pruned RNN-T loss~\cite{PrunedRNN-T} is used as the training objective function, implemented within the k2~\cite{k2} framework\footnote{https://github.com/k2-fsa/k2}. The encoder employs a 6-stack Zipformer~\cite{Zipformer} with downsampling factors of (1,2,4,8,4,2). The label decoder employs a stateless decoder~\cite{Stateless}, which consists of an embedding layer followed by a 512-dim Conv1D layer. 
A convolution subsampling module with a stride of 2 is placed to reduce the frame rate to 50 Hz before being fed into the encoder. Overall, the model has 65.5M parameters. All models are trained with 4 NVIDIA V100 32GB GPUs.

\vspace{-1em}
\subsection{Results on LibriSpeech}
\tbl{ASR-LibriSpeech-100h} shows the ASR performance of models trained on LibriSpeech 100h using different types of discrete tokens. It indicates that systems trained on tokens from WavLM and HuBERT yield improvements of 17\% and 28\% over those trained on FBank features for test-clean and test-other, respectively. The discrete tokens derived from Encodec and vq-wav2vec perform worse than FBank features. This reveals the importance of SSL models for discrete token generation and the superiority of discrete tokens from WavLM and HuBERT models on low-resource data.
We also report the results from \cite{cmu_discreteASR} to compare with the existing method. Our approach exhibits lower WERs and more substantial performance gains from self-supervised models.

\begin{table}[!h]
  \vspace{-0.7em}
  \centering
  \caption{LibriSpeech-100h WERs(\%) of FBank features and various discrete tokens.}
  \vspace{-0.6em}
  \label{tab:ASR-LibriSpeech-100h}
  \setlength{\tabcolsep}{3pt}
  \renewcommand{\arraystretch}{1.2}
  \scalebox{0.98}{
      \resizebox{1\linewidth}{!}{
        \begin{tabular}{cccccc}
        \hline
        \multirow{2}{*}{\textbf{Method}}
        & \multirow{2}{*}{\textbf{Feature}}
        & \multirow{2}{*}{\textbf{\# Units}}
        & \textbf{Bandwidth}
        & \multicolumn{2}{c}{\textbf{test}} \\
                          &              &          & (kbps) & \textbf{clean} & \textbf{other} \\ \hline
        \multirow{2}{*}{Chang et al. (2023)~\cite{cmu_discreteASR}} 
                          & FBank        & -        & 256.00 &  8.30 & 22.20 \\
                          & WavLM-Large  & 2000     &   0.55 &  5.90 & 12.80 \\ \hline
        \multirow{5}{*}{Ours} 
                          & FBank        & -        & 256.00 &  6.27 & 16.67 \\
                          & WavLM-Large  & 2000     &   0.55 &  5.22 & 11.85 \\
                          & HuBERT-Large & 2000     &   0.55 &  5.19 & 11.94 \\
                          & EnCodec      & $1024^8$ &   6.00 &  7.16 & 22.04 \\
                          & vq-wav2vec   & $320^2$  &   1.66 & 11.76 & 33.08 \\ \hline
        \end{tabular}
      }
  }
  \vspace{-0.7em}
\end{table}

\tbl{ASR-LibriSpeech-960h} presents the ASR performance of models trained on LibriSpeech 960h using different types of discrete tokens. Systems trained on tokens from WavLM and HuBERT show competitive performance against FBank-based systems. The results from \cite{cmu_discreteASR} are also given for comparison. Our approach not only achieves lower WERs but also narrows the performance gap with FBank-based systems. 
Similar to the observation on Librispeech 100h as shown in Table \ref{tab:ASR-LibriSpeech-100h}, discrete tokens from Encodec and vq-wav2vec are consistently worse than those from HuBERT and WavLM models. This could be explained by the fact that the features of HuBERT and WavLM are extracted from the top Transformer layer, which contains more semantic information.
It is also worth noting that the discrete tokens yield comparable WER results to the continuous FBank feature on the test-clean set. But in contrast, a relatively larger WER gap is observed on test-other.
One possible reason is that the discrete tokens are good at extracting semantic information under clean conditions but still incompetent for complicated acoustic conditions. 

\begin{table}[!h]
  \vspace{-0.5em}
  \centering
  \caption{LibriSpeech-960h WERs (\%) of FBank features and various discrete tokens.}
  \vspace{-0.7em}
  \label{tab:ASR-LibriSpeech-960h}
  \setlength{\tabcolsep}{3pt}
  \renewcommand{\arraystretch}{1.2}
  \scalebox{0.75}{
      \resizebox{1\linewidth}{!}{
        \begin{tabular}{cccccc}
        \hline
        \multirow{2}{*}{\textbf{Method}}
        & \multirow{2}{*}{\textbf{Feature}}
        & \multirow{2}{*}{\textbf{\# Units}}
        & \multicolumn{2}{c}{\textbf{test}} \\
                          &              &          & \textbf{clean} & \textbf{other} \\ \hline
        \multirow{2}{*}{Chang et al. (2023)~\cite{cmu_discreteASR}} 
                          & FBank        & -        &  2.60 &  6.20 \\
                          & WavLM-Large  & 2000     &  3.00 &  7.00 \\ \hline
        \multirow{5}{*}{Ours} 
                          & FBank        & -        &  2.23 &  5.15 \\
                          & WavLM-Large  & 2000     &  2.29 &  5.50 \\
                          & HuBERT-Large & 2000     &  2.29 &  5.73 \\
                          & EnCodec      & $1024^8$ &  2.57 &  6.81 \\
                          & vq-wav2vec   & $320^2$  &  3.19 &  9.44 \\ \hline
        \end{tabular}
      }
  }
\vspace{-1.7em}
\end{table}

\subsection{Results on GigaSpeech 1000h and AISHELL-1}
Previous experiments were conducted on Librispeech, which matches well with the SSL training data. To get a whole picture of the ASR performance with discrete tokens, two additional datasets are selected for ASR training and evaluation. The first data set is Gigaspeech 1000h (Gigaspeech-M), which aims to evaluate the ASR performance on a more diverse, noisy, and challenging dataset. The other test set is AISHELL-1, a Mandarin speech corpus recorded in clean condition, to assess the adaptability of discrete tokens across languages.
\tbl{ASR-GigaSpeech-1000h} shows the ASR performance of models trained on Gigaspeech 1000h and AISHELL-1. In Gigaspeech, discrete tokens extracted from WavLM are consistently superior to those of HuBERT. Discrete tokens of WavLM lag behind the continuous FBank feature by a relative WER degradation of 7\%. Despite that, the WER results achieved by the discrete tokens are quite impressive and competitive compared to previous studies \cite{GigaSpeech, LongFNT} on the Gigaspeech-M. In contrast, in the Mandarin AISHELL-1 dataset, the discrete tokens present much worse WER results than the FBank feature. Recalling that the SSL models used for extracting discrete tokens are trained in English corpus, the generalization of current discrete tokens across languages is yet to be improved.

\begin{table}[!h]
  \vspace{-0.5em}
  \centering
  \caption{GigaSpeech-1000h WERs(\%) and AISHELL-1 CERs(\%) of FBank features and various discrete tokens.}
  \vspace{-0.7em}
  \label{tab:ASR-GigaSpeech-1000h}
  \setlength{\tabcolsep}{3pt}
  \renewcommand{\arraystretch}{1.2}
  \scalebox{0.65}{
    \resizebox{1\linewidth}{!}{
    \begin{tabular}{ccccc}
    \hline
    \multirow{2}{*}{\textbf{Dataset}}
    & \multirow{2}{*}{\textbf{Feature}}
    & \multirow{2}{*}{\textbf{\# Units}}
    & \multicolumn{2}{c}{\textbf{WER}} \\ \cline{4-5} 
                                &              &      & \textbf{dev} & \textbf{test} \\ \hline
    \multirow{3}{*}{GigaSpeech} & FBank        & -    & 12.24        & 12.19         \\ \cline{2-5} 
                                & WavLM-Large  & 2000 & 13.20        & 13.05         \\
                                & HuBERT-Large & 2000 & 14.45        & 14.79         \\ \hline
    \multirow{2}{*}{AISHELL-1}  & FBank        & -    &  4.27        &  4.49         \\ \cline{2-5}
                                & WavLM-Large  & 2000 &  8.41        &  8.83         \\ \hline
    \end{tabular}
    }
  }
  \vspace{-0.4em}
\end{table}

\vspace{-1.4em}
\subsection{Ablation Study}
\tbl{ASR-Ablation-AugmentationPolicy} demonstrates the effectiveness of customized data augmentation techniques for discretized inputs. Although Gaussian noise doesn't offer immediate benefits in LibriSpeech 100h, it helps enhance the stability of the convergence curve.
\begin{table}[!h]
  \centering
  \caption{Ablation study of augmentation policy on LibriSpeech 100h using discrete tokens from WavLM.}
  \vspace{-0.5em}
  \label{tab:ASR-Ablation-AugmentationPolicy}
  \setlength{\tabcolsep}{3pt}
  \renewcommand{\arraystretch}{1.2}
  \scalebox{0.5}{
      \resizebox{1\linewidth}{!}{
        \begin{tabular}{lcc}
        \hline
        \multirow{2}{*}{\textbf{Augmentation Policy}}
        & \multicolumn{2}{c}{\textbf{test}} \\
        & \multicolumn{1}{l}{\textbf{clean}} & \multicolumn{1}{l}{\textbf{other}} \\ \hline
        Without Augmentation                 &  6.74  & 14.71 \\
        + Time Warping                       &  6.09  & 13.48 \\
        \hspace{1.0mm} + Time Masking        &  5.57  & 12.36 \\
        \hspace{2.0mm} + Embedding Masking   &  5.58  & 12.31 \\
        \hspace{3.0mm} + Frame Duplication   &  5.31  & 11.74 \\
        \hspace{4.0mm} + Gaussian Noise      &  5.24  & 11.83 \\ \hline
        \end{tabular}
      }
  }
  \vspace{-1.5em}
\end{table}

\tbl{ASR-Ablation-EmbeddingInitialization} compares the impact of different initialization methods for the embedding layer on ASR performance, indicating the incorporation of a codebook along with an additional linear layer increases parameters without notably improving performance.
\begin{table}[!h]
  \vspace{-0.5em}
  \centering
  \caption{Ablation study of embedding initialization for discrete tokens strategies on LibriSpeech 100h using discrete tokens from WavLM.}
  \vspace{-0.5em}
  \label{tab:ASR-Ablation-EmbeddingInitialization}
  \setlength{\tabcolsep}{3pt}
  \renewcommand{\arraystretch}{1.2}
  \scalebox{0.9}{
      \resizebox{1\linewidth}{!}{
        \begin{tabular}{lccc}
        \hline
        \multirow{2}{*}{\textbf{Embedding Method}}
        & \multicolumn{2}{c}{\textbf{test}} & \multirow{2}{*}{\textbf{\# Parameters}} \\
        & \multicolumn{1}{l}{\textbf{clean}} & \multicolumn{1}{l}{\textbf{other}} & \\ \hline
        Random Initialized                       &  5.24 & 11.83 & 65.7 M \\
        Codebook Initialized + Linear Projection &  5.28 & 11.71 & 67.7 M \\ \hline
        \end{tabular}
      }
  }
  \vspace{-1em}
\end{table}

\vspace{-0.5em}
\subsection{TTS Experimental Setups}
Speech resynthesis experiments are conducted to compare the speech synthesis ability between the mel-spectrogram feature and various discrete tokens. LibriTTS~\cite{LibriTTS} containing 580 hours of speech from 2306 speakers is adopted. We first filter out utterances less than 5 seconds from the train-960 subset and use the remaining for training. For CTX-vec2wav~\cite{UniCATS}, both Conformer encoders consist of 2 blocks, each with 2 attention heads and 184 attention dimensions. The mel-spectrogram encoder employs a 1D convolution with a kernel size of 5 and an output channel of 184. The HifiGAN generator has upsampled kernel size (16,10,8,4) for HuBERT, WavLM, and (16,10,4,4) for vq-wav2vec. 

In training, we randomly divide each utterance into two segments. The first segment has a duration of 2-3 seconds, which is used to extract the mel-spectrogram and acts as the acoustic context. The remaining segment is utilized to extract discrete tokens and perform vocoding. The CTX-vec2wav are optimized using the same objective as HifiGAN, supplemented by an L1 loss from the auxiliary feature adaptor. To stabilize training, we also incorporate the multi-task warmup technique proposed in \cite{VQTTS}. All models are trained with an Adam optimizer for 500k steps with the initial learning rate set to $2\times 10^{-4}$ and halved every 200k steps. A variant of Encodec, DAC \cite{DAC}, is also adopted for speech resynthesis. For the tokens from Encodec and DAC, we utilize the publicly released checkpoint to decode tokens back to the waveform without training on LibriTTS. 

Also, we trained a HifiGAN vocoder with mel-spectrogram input, which is identical to the HifiGAN generator in the vq-wav2vec variant. The mel-spectrogram feature is extracted with a shift of 10ms and a frame length of 29.0625ms. Other configurations remain consistent with those mentioned earlier. 

\vspace{-0.5em}
\subsection{Resynthesis Results on LibriTTS 960h}
To evaluate the performance of these models, we conduct resynthesis on a test set comprised of 500 randomly selected utterances\footnote{https://francissfy.github.io/universal\_discrete\_token.html} from the LibriTTS test-clean subset, which includes recordings from 37 different speakers. For the CTX-vec2wav models trained with HuBERT, WavLM, and vq-wav2vec tokens, we perform resynthesis on each utterance using a 3-second prompt from the same speaker. In the case of the HifiGAN vocoder with mel-spectrogram input, we directly perform vocoding. Finally, for the Encdec and DAC variants, we convert the discrete tokens to waveforms using the provided decoder.

The resynthesis results are assessed using MOS listening tests, where 10 participants rate naturalness and speaker similarity on a 1-5 scale. Speaker Encoder Cosine Similarity (SECS) is calculated using Resemblyzer\footnote{https://github.com/resemble-ai/Resemblyzer} as an auxiliary metric for speaker similarity.
Table \ref{tab:TTS-libritts-960h} shows the results of subjective and objective evaluations for the speech resynthesis task with various discrete tokens and mel-spectrogram features. It can be seen that all discrete tokens, except Encodec, are able to generate high-quality audio and present similar MOS and SECS performance compared to mel-spectrogram features. It is worth noting that both Encodec and DAC are not fine-tuned on the LibriTTS data as other models. 
Interestingly, the DAC gives the best resynthesis performance, while the Encodec presents the worst performance. Given the intrinsic similarity of Encodec and DAC, there is a large potential for the neural audio codec model. Discrete tokens from vq-wav2vec outperform those from HuBERT and WavLM. Recalling its inferiority in the ASR task, one possible explanation is that vq-wav2vec contains more acoustic information. Discrete tokens from HuBERT and WavLM generate high-fidelity audio with comparable quality. 

\begin{table}[!h]
\vspace{-0.5em}
\centering
\caption{Comparison of TTS performance on LibriTTS-960h among mel spectrogram and various discrete tokens.}
\vspace{-0.5em}
\label{tab:TTS-libritts-960h}
\setlength{\tabcolsep}{3pt}
\renewcommand{\arraystretch}{1.2}
\scalebox{0.95}{
    \resizebox{1\linewidth}{!}{
    \begin{tabular}{cccccc}
    \hline
    \multirow{2}{*}{\textbf{Feature}}
    & \multirow{2}{*}{\textbf{\# Units}}
    & \textbf{Bandwidth}
    & \multicolumn{2}{c}{\textbf{MOS}}
    & \multirow{2}{*}{\textbf{SECS}} \\
                    &          & (kbps) & \textbf{Naturalness} & \textbf{Similarity} &       \\ \hline
    Ground-truth    & -        & -      & 4.48                 & 4.18                & 0.843 \\
    Mel spectrogram & -        & 256.00 & 4.36                 & 4.17                & 0.834 \\ \hline
    Encodec         & $1024^8$ &   6.00 & 3.83                 & 3.85                & 0.834 \\
    DAC             & $1024^8$ &   4.00 & 4.41                 & 4.30                & 0.841 \\
    vq-wav2vec      & $320^2$  &   1.66 & 4.36                 & 4.21                & 0.842 \\
    HuBERT Large    & 2000     &   0.55 & 4.26                 & 4.18                & 0.833 \\
    WavLM Large     & 2000     &   0.55 & 4.18                 & 4.18                & 0.836 \\ \hline
    \end{tabular}
    }
}
\vspace{-1em}
\end{table}
\vspace{-0.5em}
\section{Conclusions}
With the goal of exploring the universality of speech discrete tokens across multiple speech tasks, this paper conducted a comprehensive study employing two representative speech processing tasks: speech recognition and speech synthesis.
We investigate speech discrete tokens from four leading SSL models, including vq-wav2vec, encodec, HuBERT, and WavLM.
Experimental results indicate that discrete tokens are competitive with FBank features in speech recognition and surpassed mel-spectrogram features in speech synthesis.
These empirical findings suggest that universal discrete tokens have considerable potential across diverse speech-related tasks.
It's crucial to note that this study is preliminary. While it has contributed valuable insights through its comparative analysis of various speech pre-training models and tasks, there is still ample room for refinement.
We aspire that this work may serve as a foundational stepping stone, bridging the representation of speech and text, and promoting research in the domain of cross-modal exploration.

\vfill\pagebreak
\begin{spacing}{0.5}
\bibliographystyle{IEEEbib}
\bibliography{strings,refs}
\end{spacing}

\end{document}